\documentstyle[preprint,aps,eqsecnum]{revtex}
\tighten
\input epsf.tex
\begin{document}
\draft
\title{\bf Quantum Fluctuations of Axions}
\author{{\bf  Edward W. Kolb$^{(a)}$, 
 Anupam Singh$^{(b)}$ and Mark Srednicki$^{(b)}$}}
\address
{ (a) NASA/Fermilab Astrophysics Center\\
Fermi National Accelerator Laboratory, Batavia, Illinois~~60510,  
and\\
Department of Astronomy and Astrophysics, Enrico Fermi Institute\\
The University of Chicago, Chicago, Illinois~~ 60637, U. S. A.\\
 (b) Department of Physics, University of California, 
Santa Barbara, CA 93106, U. S. A. 
 }
\maketitle
\begin{abstract}
We study the time evolution of the quantum fluctuations of the axion field
for both the QCD axion as well as axions arising in the context of
supergravity and string theories.
We explicitly keep track not only of the coherently oscillating zero
momentum mode of the axion but also of the higher non-zero momentum modes
using the full axion potential.
The full axion potential makes possible two kinds of instabilities:
spinodal instabilities and parametric resonance instabilities.
The presence of either of these instabilities can lead to a quasi-exponential
increase in the occupation of non-zero momentum modes and the build-up of the
quantum fluctuations of the axions.
If either of these becomes a significant effect then axions would no longer 
be a suitable cold dark matter candidate.
Our results confirm the conventional wisdom that these effects are not 
significant in the setting of an expanding FRW universe and hence
axions are indeed cold dark matter candidates.
\end{abstract}

\newpage

\section{\bf Introduction and Motivation}

Over the years there has been growing evidence that nonbaryonic,
cold dark matter plays an
important role in the structure and evolution of the 
universe ( see, e.g. \cite{mark,kt,jkg}).
Axions are among the most promising candidates
for the non-baryonic cold dark matter of the universe\cite{pww,as,df}.
The concept of the axion was originally introduced
and developed to solve the strong CP problem of QCD in an appealing 
and phenomenologically acceptable way\cite{qaxion}.  In these models,
axions are pseudo-Goldstone bosons of a U(1) symmetry.
After the advent of supergravity and string theories, it became clear
that particles with the properties of axions were in fact more generic,
and that in these theories there could be additional axion fields which 
may play a significant role in the history of the 
universe (see e.g. \cite{saxion}).

In this paper, we examine the time evolution of the quantum fluctuations
of the axion field as it oscillates about the minimum of its potential. 
The standard picture is that the axion oscillates coherently in its potential;
only the zero-momentum mode is important.  The original papers presenting
the axion as a dark matter candidate\cite{pww,as,df} 
considered simple estimates of 
instabilities in the axion field that could result in energy being
pumped from the zero mode into higher-momentum modes.  If such an effect
were significant, then it is possible that the energy stored in the
axion field would be largely converted to kinetic energy and subsequently
redshifted away.  According to the original estimates, these effects are
not significant.

However in recent years we have come to realize that there are two kinds of 
instabilities occurring in the time evolution of generic mode functions
that have the potential of changing this situation. Thus, either
spinodal instabilities\cite{guthpi,weinbergwu,boy1,boy2}  or 
parametric resonance instabilities\cite{branden,linde1,us1}, if they last
for a sufficiently long time, can lead to an explosive growth of quantum
fluctuations through the exponential growth of non-zero momentum modes.
Thus in the light of the recent understanding of the role of these 
instabilities in the growth of non-zero momentum modes, it is worthwhile
to re-examine the role of the quantum fluctuations of the axion.

In what follows we carefully and quantitatively study the time evolution
of both the zero and non-zero momentum modes of the axion. We are thus
able to ascertain the magnitude of the quantum fluctuations of the axion
and compare it with the value of the coherently oscillating axion field.
We do this both for the QCD axion as well as other axions that arise in the 
context of supergravity and string theories.
We show that if the axions were born, lived and died in Minkowski
space then there would in fact have been an explosive growth of 
quantum fluctuations resulting from the quasi-exponential growth of
some non-zero momentum modes of the axion field.
However, the energy density of axions is diluted by the expansion of the 
universe, which implies a decrease in the amplitude of oscillation of the
coherently oscillating zero mode.  Since it is this oscillation  
that drives the instabilities and the explosive exponential
growth of the non-zero  modes, it is clear that when the
amplitude of the zero mode falls below some critical value, the
instabilities will be shut off and there will be no further growth of
the fluctuations. The issue thus becomes one of the initial amplitudes and
timescales involved.

In our analysis  we will restrict our attention to the time evolution
of the axion zero mode and axion fluctuations. We will neglect the
couplings of the axions to other fields\cite{kaplan,mark2}.
In principle, the coupling of the axions to say  photons could result
in a parametric resonance production of photons.
In fact, parametric resonance results from the transfer of energy from the
zero mode of the axion in this case to modes of other fields.
This can only occur if the axion zero mode oscillates with a sufficiently
large amplitude for a sufficiently long time.
The central issue here is whether in the cosmological context, the
axion zero mode oscillates with a sufficiently
large amplitude for a sufficiently long time to make
parametric resonance possible.
While in our quantitative analysis we will only keep track of the
axion zero mode and axion fluctuations and neglect the couplings of
axions to other fields we expect that if the axion zero mode is not
able to excite the non-zero modes of the axion then it will also
not be able to excite the modes of other fields. If in fact we did
find that the non-zero momentum modes of axions were getting excited
by parametric resonance then we would have to keep track of the modes
of other fields.

We now turn to a quantitative analysis of the problem to determine whether
there is enough time to build up the fluctuations significantly.
In the next section we will describe and layout the equations that determine 
the time evolutions of the axion field and its fluctuations.

An important point is that
we will keep the terms in the potential which are
non-linear in the mean field $\phi/f_a$; that is, we will keep the full
expression $\cos\left(\phi/f_a \right)$ instead of just the terms
up to quadratic order in the mean field $\phi^2$. This is necessary
since the initial value of the mean field can be of order one,
and so neglecting the higher order terms in the potential is unwarranted.
However, the main issue we wish to address in this paper is whether or not
the fluctuations in the axion field ever become large due to the presence
of instabilities in the evolution equations for the mode functions. 
For this purpose it is sufficient to start off with the initial 
fluctuations small, and then keep only the lowest order term in the 
fluctuations of the axion $\langle \eta^2 \rangle/f_a^2$, where
$\eta$ represents the fluctuations in the axion field about its mean
value $\phi$.  If in following the time evolution 
we were to find that  
$\langle \eta^2 \rangle/f_a^2$ did become of order one, then we would be
required to deal with the non-linear terms in the fluctuations in
some way, such as the Hartree approximation( see e.g. \cite{frw}). 
However, as we shall see, the fluctuations never become large 
in a cosmological setting,  and so we will not be required
to use the Hartree approximation.  However, in order
to arrive at this conclusion, it is nevertheless necessary to keep the
non-linear terms in the mean field $\phi/f_a$ in the axion potential.
Further, we also consider the explicit time dependence of the axion
potential in a cosmological setting due to the temperature dependence
of the instanton effects that give rise to the axion potential.
These issues are discussed in greater detail in section 2.

In section 3 we will study and analyze the solution to these equations for
(i) the QCD axion and 
(ii) other axions arising from supergravity and string theories. 
In both cases we will compare the evolution in Minkowski space with
the behavior in an expanding FRW universe to gain insight into the
relative roles of instabilities and the expansion of the universe.
Finally we will conclude by stating the implications of our results and 
place things in perspective.

\section{\bf Evolution Equations}

The derivation of the appropriate evolution equations\cite{ctp,hu} 
has been intensively studied during the past few years by a number of groups
\cite{branden,linde1,us1,numanal,largen,jackiwetal,linde2,klr,prestable,krt,son,kaiser,yoshi,usfrw}. 
Here we will
summarize the key formulae along the lines
presented by Boyanovsky, de Vega and Holman\cite{frw}. 

In a spatially flat FRW cosmology, the metric is
\begin{equation}
ds^2 = dt^2-a^2(t)d\vec{x}^2 \ .
\end{equation}
The action and Lagrangian density for the axion field $\Phi$ are given by
\begin{eqnarray}
S         & =  & \int d^4x {\cal{L}} \label{action} \\
{\cal{L}} & =  & a^3(t)\left[\frac{1}{2}\dot{\Phi}^2(\vec{x},t)-\frac{1}{2}
\frac{(\vec{\nabla}\Phi(\vec{x},t))^2}{a(t)^2}-V(\Phi(\vec{x},t))\right]
 \label{lagrangian} \\
V(\Phi , T)   & =  & \Lambda^4(T) \left[1 - \cos\left(
\frac{ \Phi(\vec{x},t) }{f_a} \right) \right] \label{potential} \ ,
\end{eqnarray}
where $f_a$ is the axion decay constant and $T$ is the temperature, and
$\Lambda(T)$ is related to $f_a$ and the axion mass $m_a(T)$ 
via $\Lambda^2(T) = f_a m_a(T)$.
The temperature dependence of $m_a$ and $\Lambda$
is due to the temperature dependence of the instanton effects
which give rise to the axion potential\cite{gpy,kt10}.

For reasons explained in section 3, we will concentrate for the most part
on the time and temperature dependence of the QCD axion mass, and in the case
of axions arising in the context of supergravity and string theories,
we will restrict our attention to the case of a time independent potential.

For our purposes the temperature dependence of the QCD axion mass can 
be captured by the expression\cite{gpy,kt10}:
\begin{equation}
m_a(T) = 0.1\,m_{a0}\left( \frac{\Lambda_{QCD}}{T} \right)^{3.7}\,,
\label{masst}
\end{equation}
where $m_{a0}$ is the zero-temperature axion mass.
It should of course be noted that this form of the above expression 
is only valid for $T >~ \Lambda_{QCD}$.

We will be interested in two kinds of axions:  the standard QCD axion for
which $\Lambda_{QCD} \sim 200\,$MeV and 
$f_a \sim 10^{12}\,$GeV,
and axions that can arise in the context of supergravity and string theories,
for which we will take $\Lambda \sim 10^{16}\,$GeV and $f_a \sim M_{Pl}
\sim 10^{19}\,$GeV (although the parameters are much less constrained
in this case).  Axions with these parameters
have  been considered earlier in the context of natural 
inflation\cite{natural}. 
We will now consider computing the magnitude of the fluctuations of the
axion field compared to the amplitude of the coherently oscillating zero
momentum mode of the axion for general values of $\Lambda$ and $f_a$.

The canonical momentum conjugate to $\Phi$ is
\begin{equation}
\Pi(\vec{x},t) = a^3(t)\dot{\Phi}(\vec{x},t) \label{canonicalmomentum} \ ,
\end{equation}
and the Hamiltonian becomes
\begin{equation}
H(t) = \int d^3x \left\{ \frac{\Pi^2}{2a^3(t)}+
\frac{a(t)}{2}(\vec{\nabla}\Phi)^2+
a^3(t) V(\Phi) \right\} \label{hamiltonian} \ .
\end{equation}
In the Schr\"{o}dinger representation (at an arbitrary fixed time
$t_o$), the canonical momentum is represented
as
\[ \Pi(\vec{x}) = -i \frac{\delta}{\delta \Phi(\vec{x})} \ .\]  
Wave functionals obey the time dependent functional Schr\"{o}dinger equation
\begin{equation}
i \frac{\partial \Psi[\Phi,t]}{\partial t} = H \Psi[\Phi,t] \ . 
 \label{schroedinger}
\end{equation}

For the systems we'll be interested in it is convenient to
work with a functional density matrix $\hat{\rho}$ with matrix elements in the
Schr\"{o}dinger representation
$\rho[\Phi(\vec{.}), \tilde{\Phi}(\vec{.});t]$. 
 Normalizing the density matrix such that $Tr\hat{\rho}=1$, the
``order parameter'' is defined as
\begin{equation}
\phi(t) = \frac{1}{\Omega}\int d^3x \langle \Phi(\vec{x},t) \rangle =
\frac{1}{\Omega}\int d^3x  Tr\hat{\rho}(t)\Phi(\vec{x})  \ ,
\label{orderparameter}
\end{equation}
where $\Omega$ is the comoving volume, and the scale factors cancel
between the numerator (in the integral) and the denominator. Note that we have
used the fact that the field operator does not evolve in time in this picture.
In this paper we will use the terms ``order parameter'', ``mean value of the
field'' and ``zero momentum mode of the field'' interchangeably to refer to
quantity defined above.
The evolution equations for the order parameter are 
\begin{eqnarray}
\frac{d \phi(t)}{dt} & = & \frac{1}{a^3(t)\Omega}\int d^3x
  \left\langle \Pi(\vec{x},t) \right\rangle
 =\frac{1}{a^3(t)\Omega}\int d^3x  Tr\hat{\rho}(t)\Pi(\vec{x}) = \frac{\pi(t)}
{a^3(t)} \label{fidot} \\
\frac{d \pi(t)}{dt}     & = & -\frac{1}{\Omega}\int d^3x a^3(t) \left\langle
 \frac{\delta V(\Phi)}{\delta \Phi(\vec{x})} \right\rangle \label{pidot} \ .
\end{eqnarray}
It is now convenient to write the field in the {Schr\"{o}dinger} picture as
\begin{eqnarray}
\Phi(\vec{x})   & = & \phi(t)+\eta(\vec{x},t) \label{split} \\
\langle \eta(\vec{x},t) \rangle
     & = & 0 \label{doteta} \ .
\end{eqnarray}
Expanding the right hand side of (\ref{pidot}) in powers of $\eta(\vec{x},t)$
we find the effective equation of motion for the order parameter:
\begin{equation}
\frac{d^2 \phi(t)}{dt^2}+3 \frac{\dot{a}(t)}{a(t)}
\frac{d \phi(t)}{dt}+V'(\phi(t))+\frac{V'''(\phi(t))}{2 \Omega}\int d^3x
\langle \eta^2(\vec{x},t)\rangle+O(\eta^4)
=0  \ . 
\label{effequation}
\end{equation}
where primes stand for derivatives with respect to $\phi$.  
For our case,
with $V(\Phi)$ given by (\ref{potential}), we find
\begin{equation}
\frac{d^2 \phi(t)}{dt^2}+3 \frac{\dot{a}(t)}{a(t)}
\frac{d \phi(t)}{dt}+\frac{\Lambda^4(T)}{f_a} \sin\left(\frac{\phi}{f_a}\right)
\left[ 1 - \frac{\langle \eta^2 \rangle}{2 f_a^2} \right] +O(\eta^4)
=0 \ .
\label{lordereffequation}
\end{equation}
It is legitimate to neglect the $O(\eta^4)$ and higher terms in this equation
as long as $\langle\eta^{2n}\rangle/f_a^{2n}$ remains small.  We will
assume that this is the case provided
$\langle \eta^2 \rangle/f_a^2$ remains small.
Of course, if we were to find $\langle \eta^2 \rangle/f_a^2$ of order
$ 1$ at some time, then
beyond that time it would not be legitimate to neglect the higher order
terms, and we would need to keep track of them
through some non-perturbative technique 
such as the Hartree approximation. 
The first issue to address is if and when $\langle \eta^2 \rangle/f_a^2$
ever does become of order $ 1$,
and this we can do while neglecting the $O(\eta^4)$ terms.
To do this it is sufficient to consider only the lowest order terms in
$\langle \eta^2 \rangle/f_a^2$.

In order to follow the time evolution of the fluctuations
$\langle \eta^2 \rangle$ it is 
convenient to introduce  mode functions
$\varphi_k(t)$ which obey a simple evolution equations:
\begin{equation}
\frac{d^2 \varphi_k}{dt^2} + 3\frac{\dot{a}}{a}\frac{d\varphi_k}{dt}+
\left[\frac{\vec{k}^2}{a^2}+\frac{\Lambda^4(T)}{f_a^2} 
\cos\left(\frac{\phi}{f_a}\right)
\left[ 1 - \frac{\langle \eta^2 \rangle}{2 f_a^2} \right] \right]\varphi_k =0 \ ,
\label{fievolution}
\end{equation}
where we have dropped the $O(\eta^4)$ and higher terms.
In terms of the functions $\varphi_k(t)$ the
initial conditions are taken to be
\begin{eqnarray}
\varphi_k(t_o)               & = & [a^3(t_o)W_k(t_o)]^{-1/2}
 \label{bounfi} \\
\dot{\varphi}_k(t)\mid_{t_o} & = &
 i [a^{-3}(t_o)W_k(t_o)]^{1/2}
\label{bounfidot}
\end{eqnarray}
where $W_k(t_o) = [a^{-2}(t_o)k^2+m_0^2]^{1/2}$.
This initial condition corresponds to taking an initial gaussian wave-packet 
for the field $\Phi$ with a width determined by the parameter $m_0$ which 
has the dimensions of mass. 
The smaller the parameter $m_0$ the more sharply peaked the initial
distribution is around $k = 0$.
Since we are interested in studying the process of building up the occupation
of higher non-zero momentum modes starting off with essentially only the
zero momentum mode we are interested in the case where $m_0$ 
is much less than $m_a$.
We have also varied $m_0$ and checked that the rate of the build-up
of fluctuations is insensitive to the choice of $m_0$. This is actually
so because when the non-zero momentum modes do grow they do so at an
exponential rate and hence any differences in the initial occupation
of modes quickly becomes insignificant.

The equal time two-point function for the fluctuations which we have been
denoting by  $\langle \eta^2 \rangle$ can then be expressed as: 
\begin{equation}
\langle \eta^2 \rangle = \frac{1}{2}\int \frac{d^3k}{(2\pi)^3}
\frac{\mid \varphi_k(t) \mid^2}{2}. \label{fluctuations}
\end{equation}

At this point, it is worth discussing the momentum cutoff in the integral
above. The potential for the axion is non-renormalizable and the theory
we have described above should be considered only as a low energy
effective field theory. Having said that we still have to address the
issue of the cutoff in the momentum integral. First it should be noted
that the instabilities in the mode functions can occur for momenta
$k$ less than or order of $m_a$. Thus for instance for the QCD axion, 
if we choose the momentum cutoff
$k_{cutoff}$ such that $ m_a << k_{cutoff} << \Lambda_{QCD}$ then
we will certainly capture any instabilities in the growth of the mode
functions. In particular, we have taken the momentum cutoff $\sim 10^2 m_a$
and verified that our results are insensitive to the cutoff. Note
that when the instabilities in the mode evolution equation actually
kicks in (as happens for instance in Minkowski space) the mode functions
which can undergo a quasi-exponential expansion are well within
our momentum cutoff.

Having described the formalism and arrived at the evolution equations for 
the axion fields of interest for us we are now in a position to study the
solutions to these evolution equations. We turn to this in the following
section.
At this point we point out that there are some aspects of a similar
problem specifically in the context of natural inflation that are
currently being investigated\cite{dpr}.

\section{Solutions to the evolution equations and analysis.}

To analyze the solutions to the evolution equations we have written
down in the previous section it is convenient to rescale variables
into dimensionless ones in the following way.

For a radiation dominated (RD) universe,
the temperature dependence of the axion potential can be translated
into an explicit time dependence by using\cite{kt10}
\begin{eqnarray}
H(t) &=& {1\over 2t} \\
     &=&  \frac{5}{3} g_*^{1/2}(T)\,\frac{T^2}{m_{pl}} 
\end{eqnarray}
where $H(t)$ is the Hubble parameter as a function of time and
$g_*(T)$ is the number of effective degrees of freedom at temperature $T$.
The oscillations of the axion field in its potential begin approximately
at a temperature $T_1$ such that $3H(T_1) = m_a(T_1)$;
for the QCD axion, $T_1 \simeq 0.9\,GeV$.
It turns out to be convenient to write out the evolution equations in terms
of dimensionless quantities defined in units of $m_a(T_1)$ 
in the following way:
\begin{equation}
\xi  =  \frac{\phi}{f_a}\, , \quad
\tau   =   m_a(T_1) t\, , \quad
\langle \chi^2 \rangle   =   \frac{\langle \eta^2 \rangle}{2 f_a^2}\, , \quad
\eta_k   =  \phi_k \sqrt{m_a(T_1)}\, , \quad
q   = \frac{k}{m_a(T_1)} \, . 
\label{scale}
\end{equation}
We further introduce the dimensionless coupling
\begin{equation}
g =  \frac{1}{8 \pi^2}  \left( \frac{m_a(T_1)}{f_a} \right) ^2 \, .
\label{g}
\end{equation}
In terms of these dimensionless variables, the rescaled equations are
\begin{eqnarray}
&&  \frac{d^2 \xi(\tau)}{d \tau^2}+3 \frac{\dot{a}(\tau)}{a(\tau)}
\frac{d \xi(\tau)}{d\tau}+ (0.1) \tau^{3.7} \sin \xi
\left[ 1 - \langle \chi^2 \rangle \right]
=0  \, ,
\label{ev1} \\
&&  \left[\frac{d^2}{d\tau^2}+3 \frac{\dot{a}(\tau)}{a(\tau)}\frac{d}{d\tau}+
\frac{q^2}{a^2(\tau)}+ (0.1) \tau^{3.7} \cos \xi
\left[ 1 - \langle \chi^2 \rangle \right] \right]
 \eta_k(\tau) =0 \, , 
\label{ev2} \\
&& \langle \chi^2 \rangle = g
\int q^2 dq \frac{|\eta_k(t)|^2}{2} \, . 
\label{ev3}
\end{eqnarray}

We will study axions and their fluctuations in two different space-time
settings.  We do this in order to compare the behavior and gain insight
into the physics of the build-up of fluctuations. 
Thus we will study axions in a 
radiation dominated (RD) expanding universe, and in a static Minkowski
space-time. Both of these situations can be captured by parameterizing
the time evolution of the scale factor as
\begin{equation}
a(\tau)=a_0 \tau^{n} .
\end{equation}
We will take $a_0 = 1$ and consider the two cases $n = 1/2$
and  $n = 0$ corresponding to an RD Universe and
Minkowski space-time respectively. 
For each space-time setting we study two different
kinds of axions: the QCD axion and axions that arise in the context of
supergravity and string theories.  These are parameterized by their
different values of $g$.

Our goal here is to study the question of whether instabilities in the
mode equations can lead to a significant build up of quantum fluctuations
of axions. We are interested in  two kinds of axions:  the standard QCD axion 
and axions that can arise in the context of supergravity and string theories. 
We will concentrate most of our effort here in the study of QCD axions
for 2 reasons:
(i) the parameters for the QCD axion are more tightly constrained by
phenomenological, astrophysical and cosmological considerations and
(ii) the potential implications of the quantum fluctuations of QCD axions
has a more pressing relevance for cosmological issues.
Thus we will study the time evolution of the QCD axion taking into
account both the expansion of the universe as well as the explicit 
time dependence of the axion potential in a cosmological context.
We will see that the time dependence of the axion potential only
helps in further suppressing the growth of axion fluctuations.
Thus if the axion fluctuations are unable to grow in a time independent
potential, they will not grow in a potential in which mass increases as
a function of time. Hence for the axions arising in the context of
supergravity and string theories we will only investigate the growth of
axion fluctuations in a time independent potential. We will see that
for these axions too the fluctuations in an expanding universe do not
grow even in a time independent potential.

These two different kinds of axions have
very different values of $\Lambda$ and $f_a$ which enter into the quantity
$g$ which arose in the way we rescaled variables into a dimensionless form.
Thus for the QCD axion we have $g = 3 \times 10^{-60}$, 
and for string and supergravity axions we have $g =  10^{-14}$.
In this latter case, as we have already explained,
we will restrict our attention to the case of a time independent potential.
The relevant equations are the same as those of 
eqs.~(\ref{scale}--\ref{ev3}), but with $m_a(T_1)$ replaced by
$m_{a0}$; also, the factors of $0.1\,\tau^{3.7}$ are absent
in eqs.~(\ref{ev1}--\ref{ev3}).

Having done all the above things,
the solutions to the evolution equations can be displayed in terms
of plots of quantities as a function of time. We have organized these
into four figures: 
Figure 1 has plots representative of the QCD axion with a time
{\em dependent} mass in a RD universe 
$\left(n = 1/2, g = 3 \times 10^{-60} \right)$;
Figure 2 has plots representative of the QCD axion with a time
{\em independent} mass in a RD universe 
$\left(n = 1/2, g = 3 \times 10^{-60} \right)$;
Figure 3 has plots representative of supergravity and string theory axions 
in a RD universe 
$\left(n = 1/2 , g = 10^{-14} \right)$;
Figure 4 has plots representative of supergravity and string theory axions 
in a  Minkowski space-time
$\left(n = 0 , g = 10^{-14} \right)$.

For each of these cases above we plot the time evolution of the
axion field zero momentum mode ($\xi$); and the
logarithm of the axion fluctuations  ($ \log [\langle \chi^2 \rangle] $).

We wish to emphasize that the purpose of figures 2 and 4 is merely to
provide a clearer picture of the different roles played by the time
dependence of the axion mass and of a static Minkowski space versus an
expanding FRW universe respectively.
Thus figures 1 and 2 differ only in the fact that whereas figure 1 displays
the evolution with the  time  dependent axion mass in the cosmological
setting, figure 2 has the evolution with everything the same except
that the axion mass is frozen at the value at which the axions would
start oscillating.
Thus, by comparing figures 1a and 2a it is clear that the effect of the
increasing axion mass is to increase the frequency of oscillations and
to cause a more rapid dying out of the amplitude of oscillations.
Further, by comparing figures 1b and 2b it is clear that the fluctuations
die more rapidly in the presence of a time dependent axion mass.
Finally, figures 3 and 4 highlight the difference between the
Minkowski space and the FRW universe behavior and importance of 
axion fluctuations. In particular, it shows that with a time independent
axion mass, axion fluctuations would
quickly become significant in Minkowski space whereas they would not be
significant in an expanding universe setting.
It should also be pointed out that in fact the qualitative behavior
for both the axions arising in the context of string and supergravity 
theories and the QCD axion are similar, however there are quantitative
differences because the parameters are very different in the different cases.

\section{\bf Conclusions}

In this paper we examined the time evolution of the axion field and
its quantum fluctuations. This was done by keeping track not only
of the zero momentum mode of the axion field but also of the higher
non-zero momentum modes of the axion. We studied different kinds of axions:
the axion required to solve the strong CP problem in the context of QCD, as
well as other axions that would arise in the context of supergravity and
string theories. Further, we studied the dynamics in two different space-time
settings: in Minkowski space as well as in an expanding radiation dominated
FRW universe.

This study was motivated by the recent progress in understanding the
important role of phenomena such as spinodal instabilities and
parametric resonance instabilities that can lead to a rapid build-up
of fluctuations. Indeed, we found that in Minkowski space axion
fluctuations rapidly build up in a quasi-exponential manner. In the
expanding universe there are a few additional factors which we
understand through our quantitative analysis. First, we note that
because the axion is a pseudo-Goldstone boson of a compact $U(1)$
symmetry, the potential for the axion is periodic and hence the
displacement from the minimum is bounded. This limits the initial
amplitude of coherent oscillations of the axion. Further, since the
instabilities in the higher momentum modes are driven by the coherent
oscillations of the zero momentum mode it is clear that if the amplitude of
oscillations drop below some threshold the instabilities will not get a
chance to really take off. The result of our analysis is that in fact
in an expanding universe quantum fluctuations of the axion do not become
significant.

Indeed this result preserves a central piece of axion lore.
If the higher non-zero momentum modes had been significantly occupied
and quantum fluctuations had not been negligible then it would not
have been accurate to think of axions as oscillating coherently and
axions would not have been a suitable cold dark matter candidate.

\acknowledgements

We thank Rich Holman for helpful discussions.
A.S. and M.S. were supported by the National Science Foundation 
grant PHY-91-16964.
EWK was supported by the DOE and NASA under 
Grant NAG5-2788.

\begin{figure}
\centerline{ \epsfxsize=420pt  \epsfbox{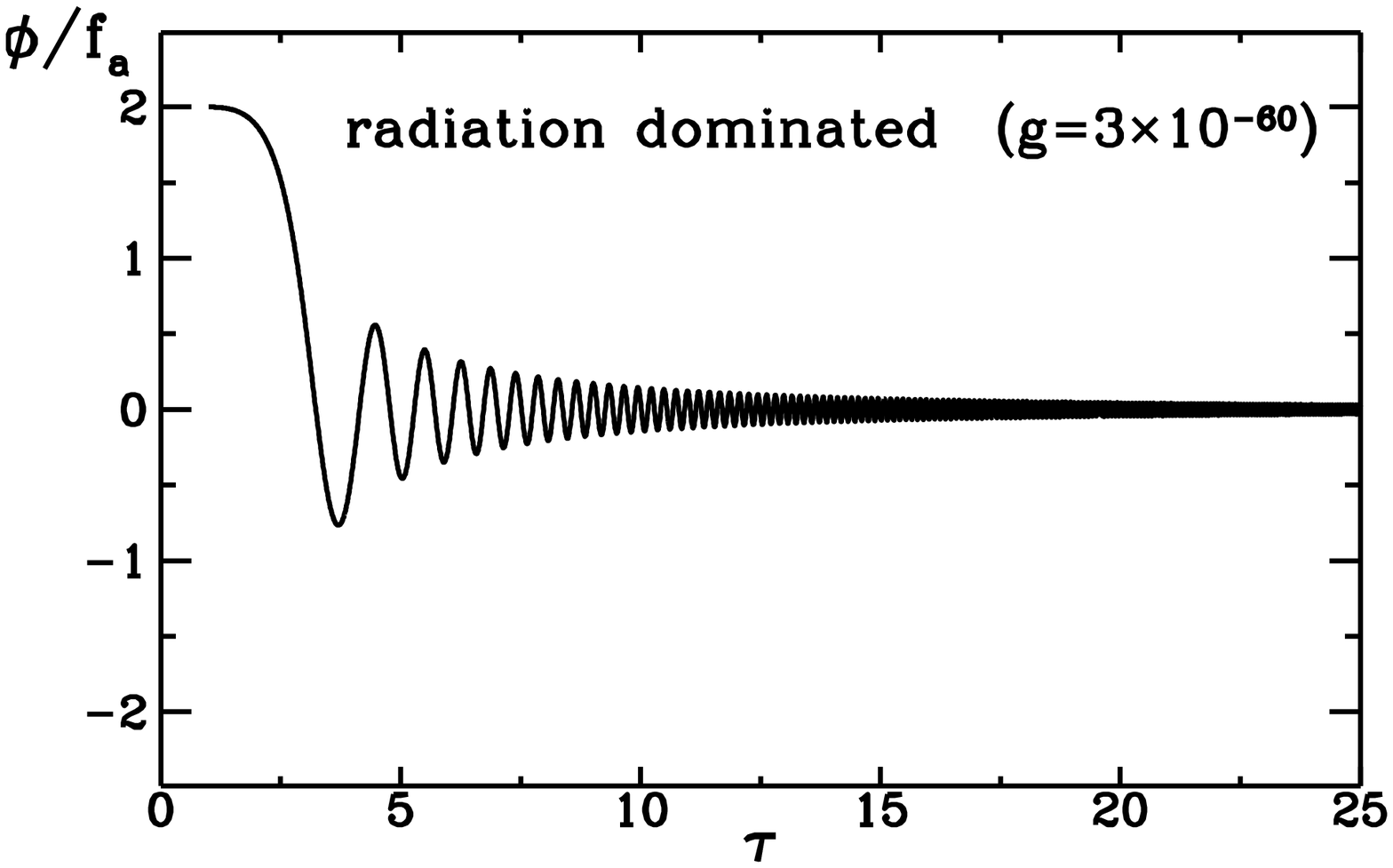} }
\footnotesize{Fig.\ 1a: QCD axion field with time dependent axion mass 
in a  radiation-dominated universe.}\\
\centerline{ \epsfxsize=420pt  \epsfbox{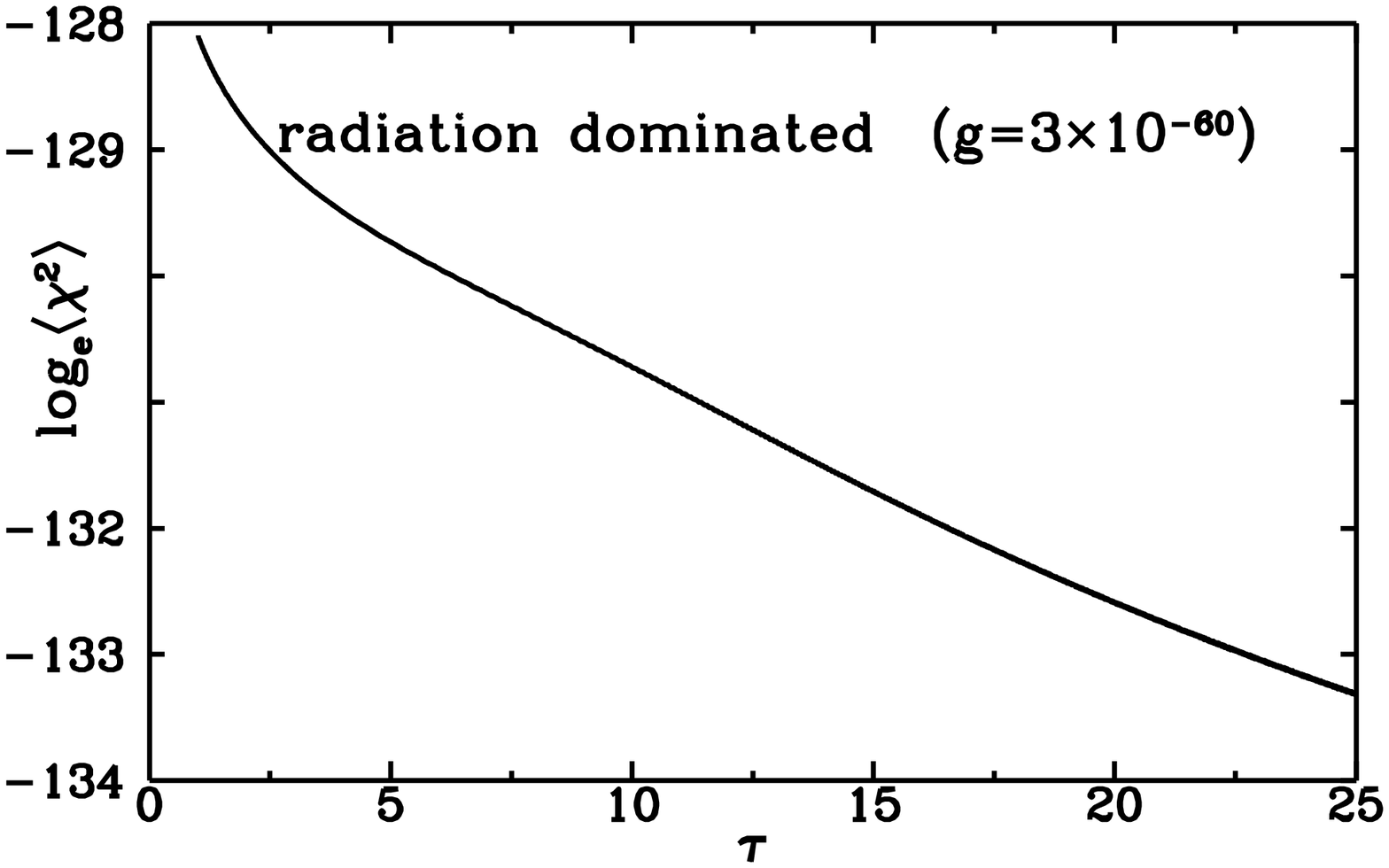} }
\footnotesize{Fig.\ 1b: QCD axion fluctuations with time dependent axion 
mass in a  radiation-dominated universe.}\\
\end{figure}

\newpage
\begin{figure}
\centerline{ \epsfxsize=420pt  \epsfbox{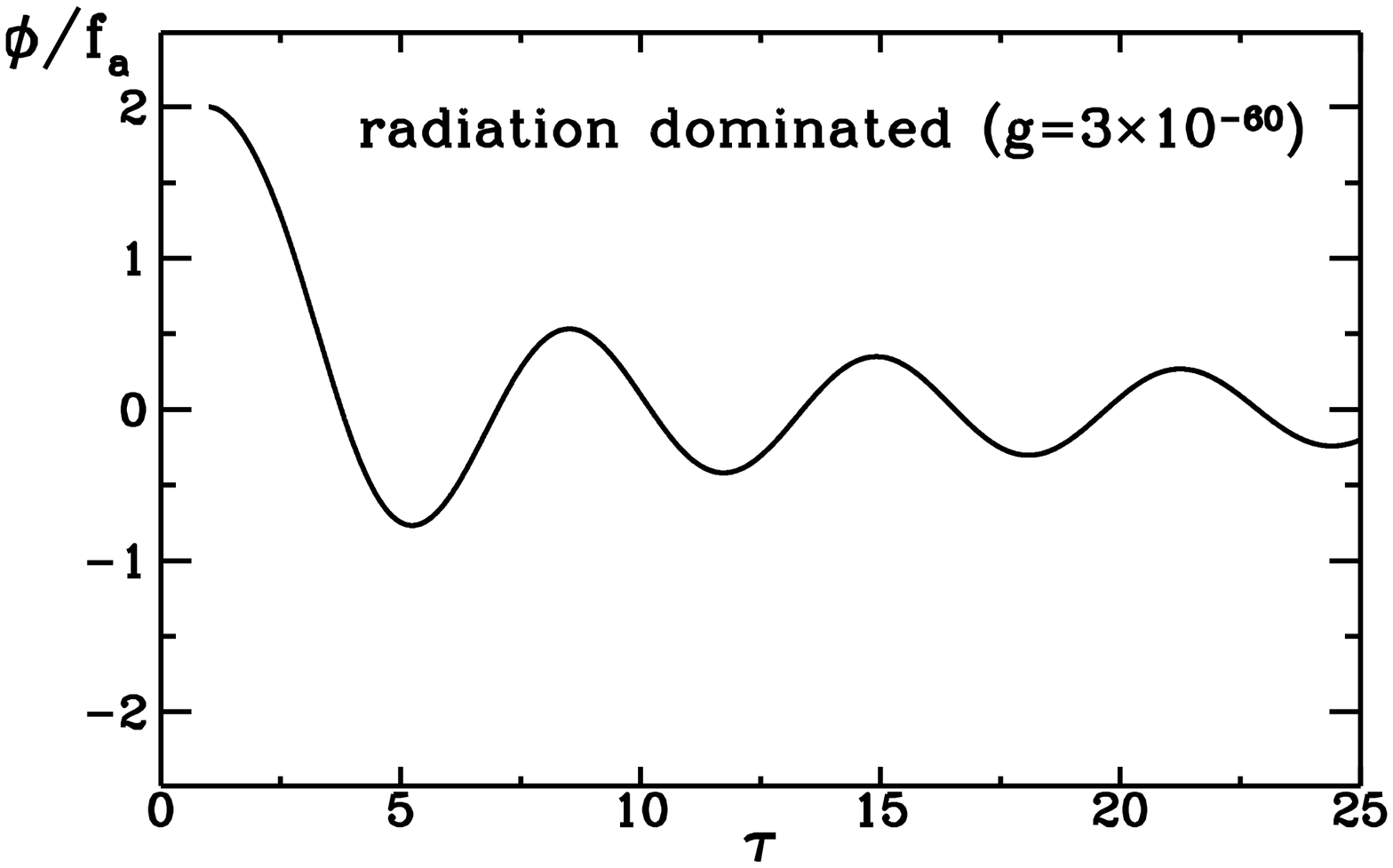} }
\footnotesize{Fig.\ 2a: QCD axion field with time independent axion mass 
in a  radiation-dominated universe.}\\
\centerline{ \epsfxsize=420pt  \epsfbox{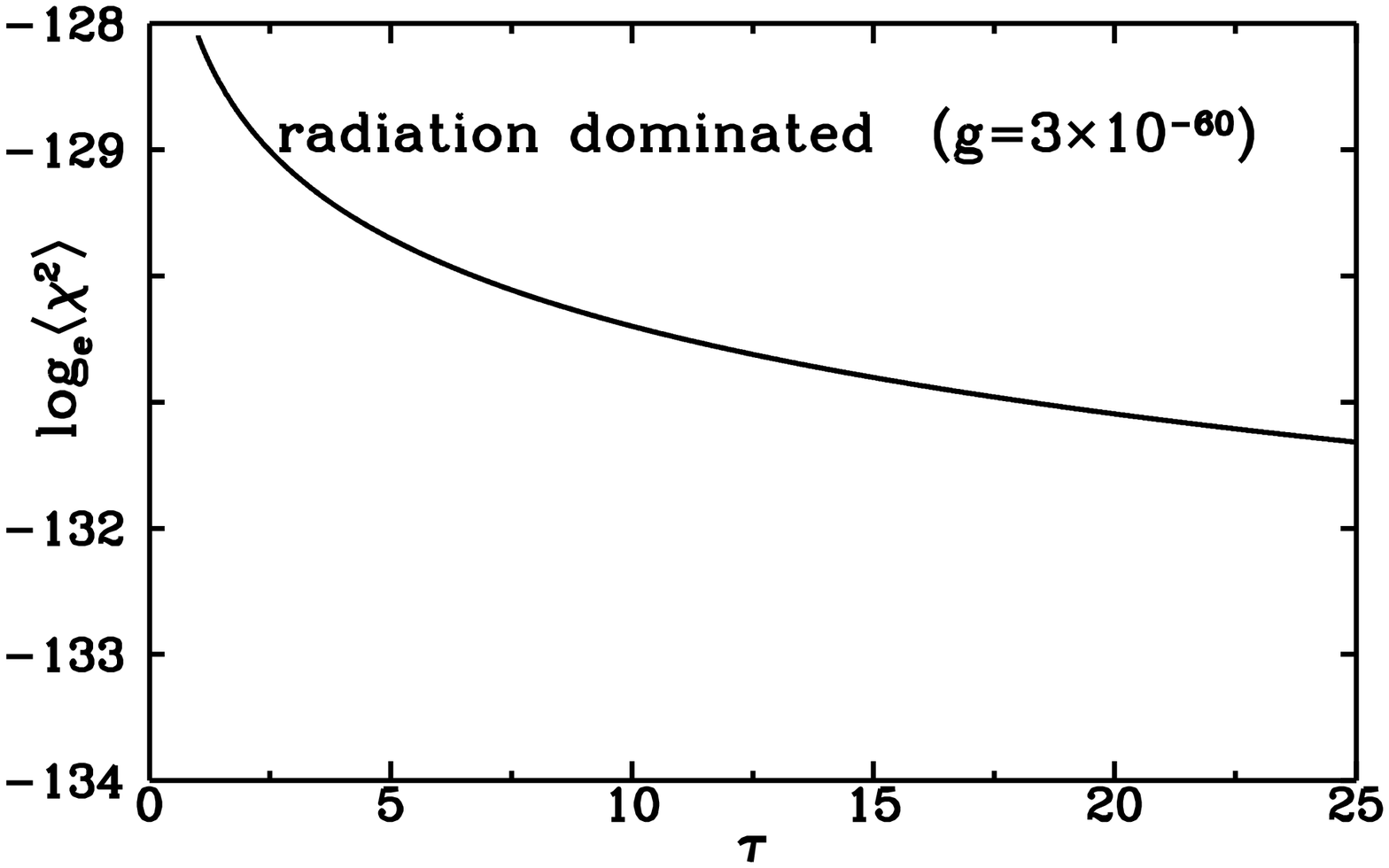} }
\footnotesize{Fig.\ 2b: QCD axion fluctuations with time independent axion 
mass in a  radiation-dominated universe.}\\
\end{figure}

\newpage
\begin{figure}
\centerline{ \epsfxsize=420pt  \epsfbox{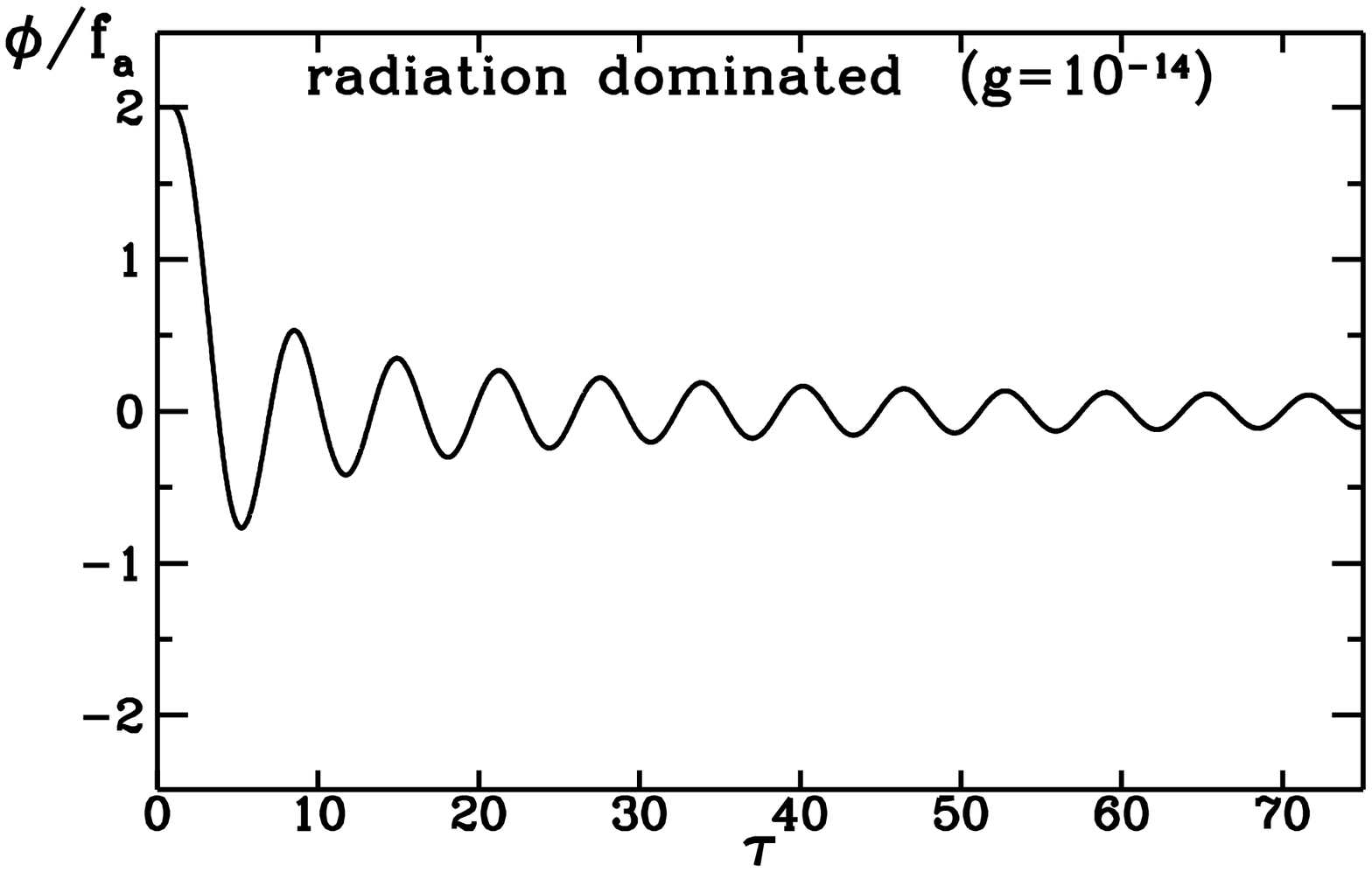} }
\footnotesize{Fig.\ 3a.  Time evolution of the supergravity and string theory 
axion zero momentum mode in a radiation-dominated universe.}\\
\centerline{ \epsfxsize=420pt  \epsfbox{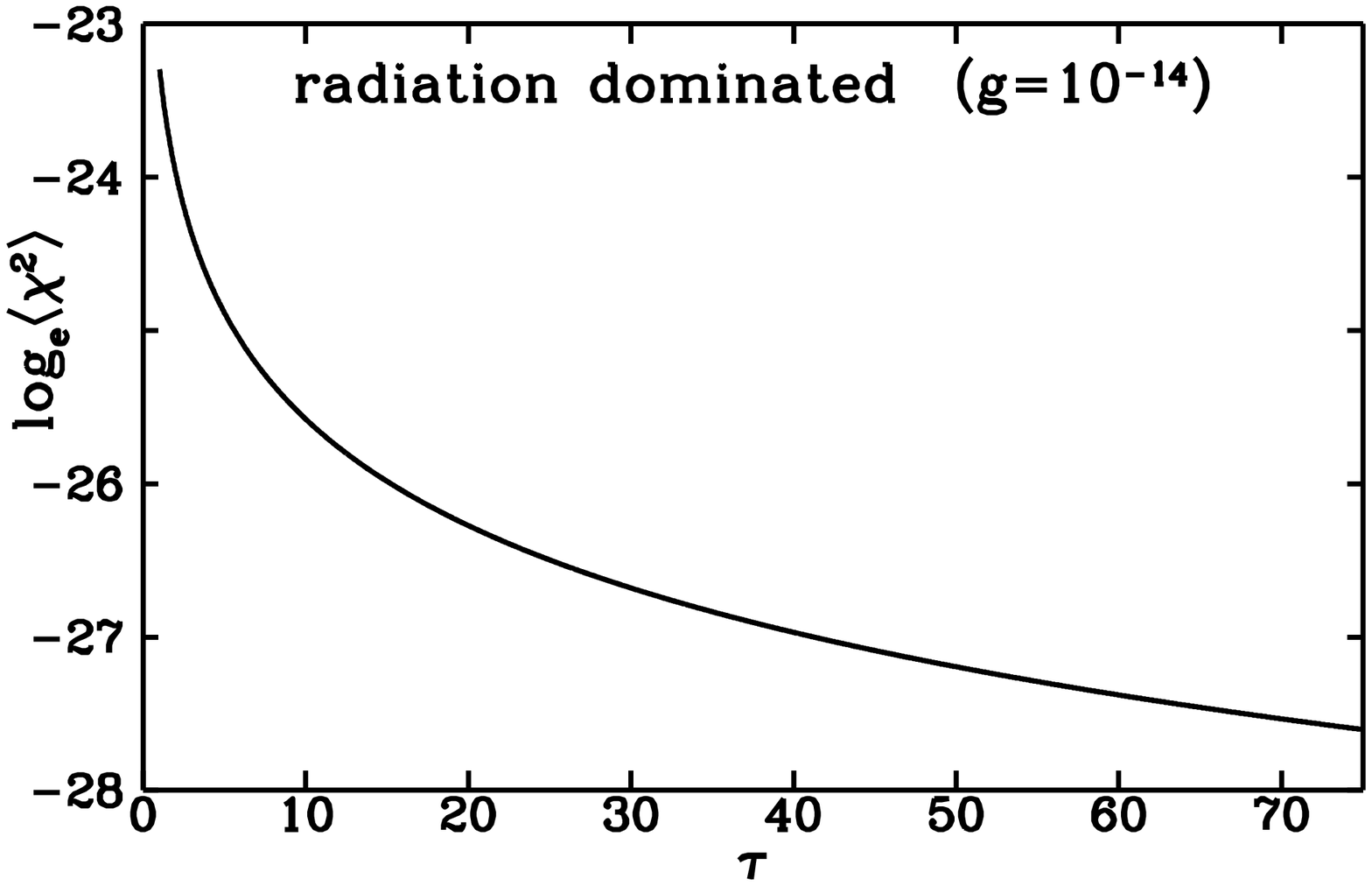} }
\footnotesize{Fig.\ 3b:  Time evolution of the supergravity and 
string theory  axion fluctuations in a radiation-dominated universe.}\\
\end{figure}

\newpage
\begin{figure}
\centerline{ \epsfxsize=420pt  \epsfbox{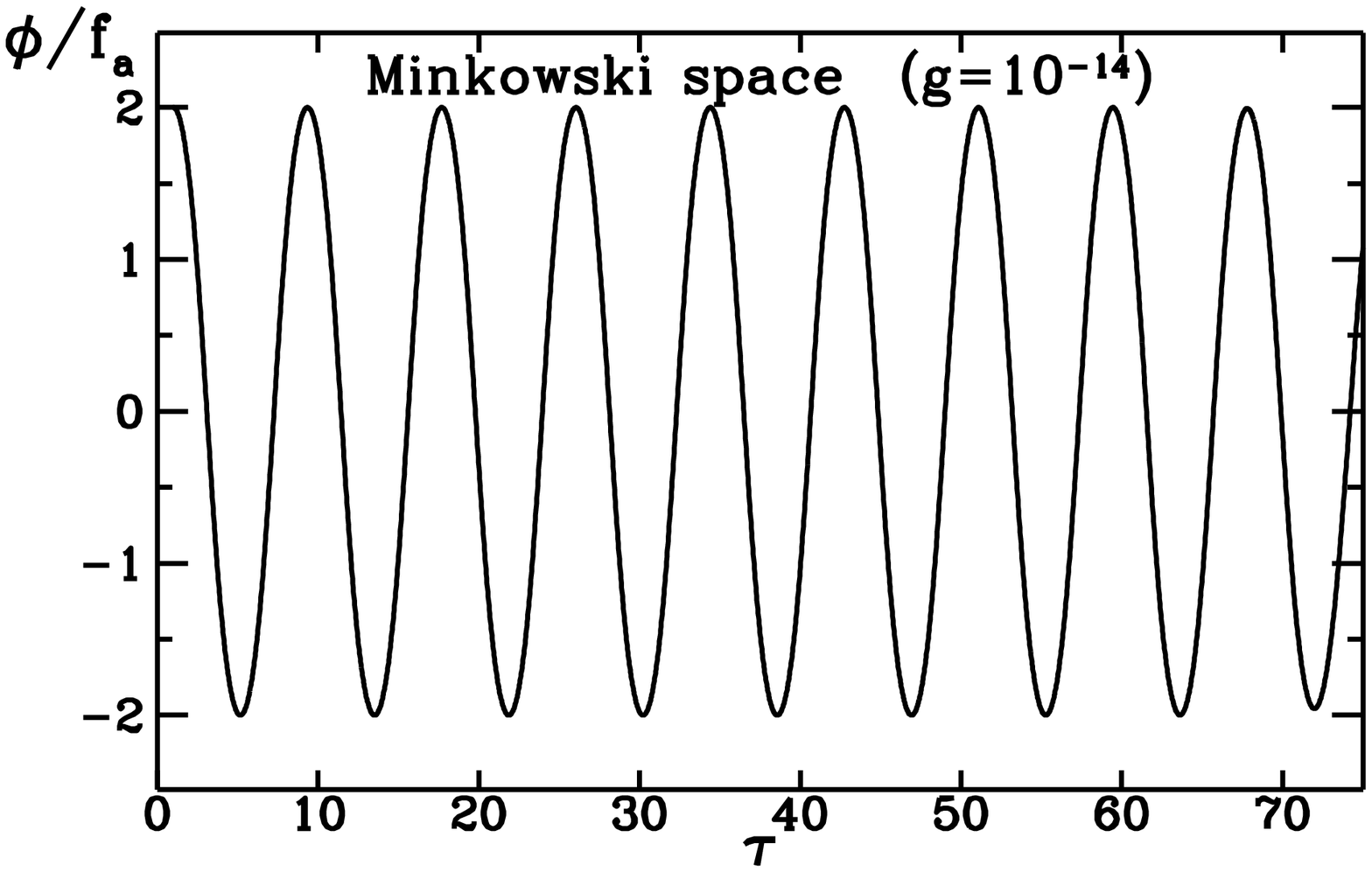} }
\footnotesize{Fig.\ 4a:  Time evolution of the supergravity and string theory 
axion zero momentum mode in Minkowski space.}\\
\centerline{ \epsfxsize=420pt  \epsfbox{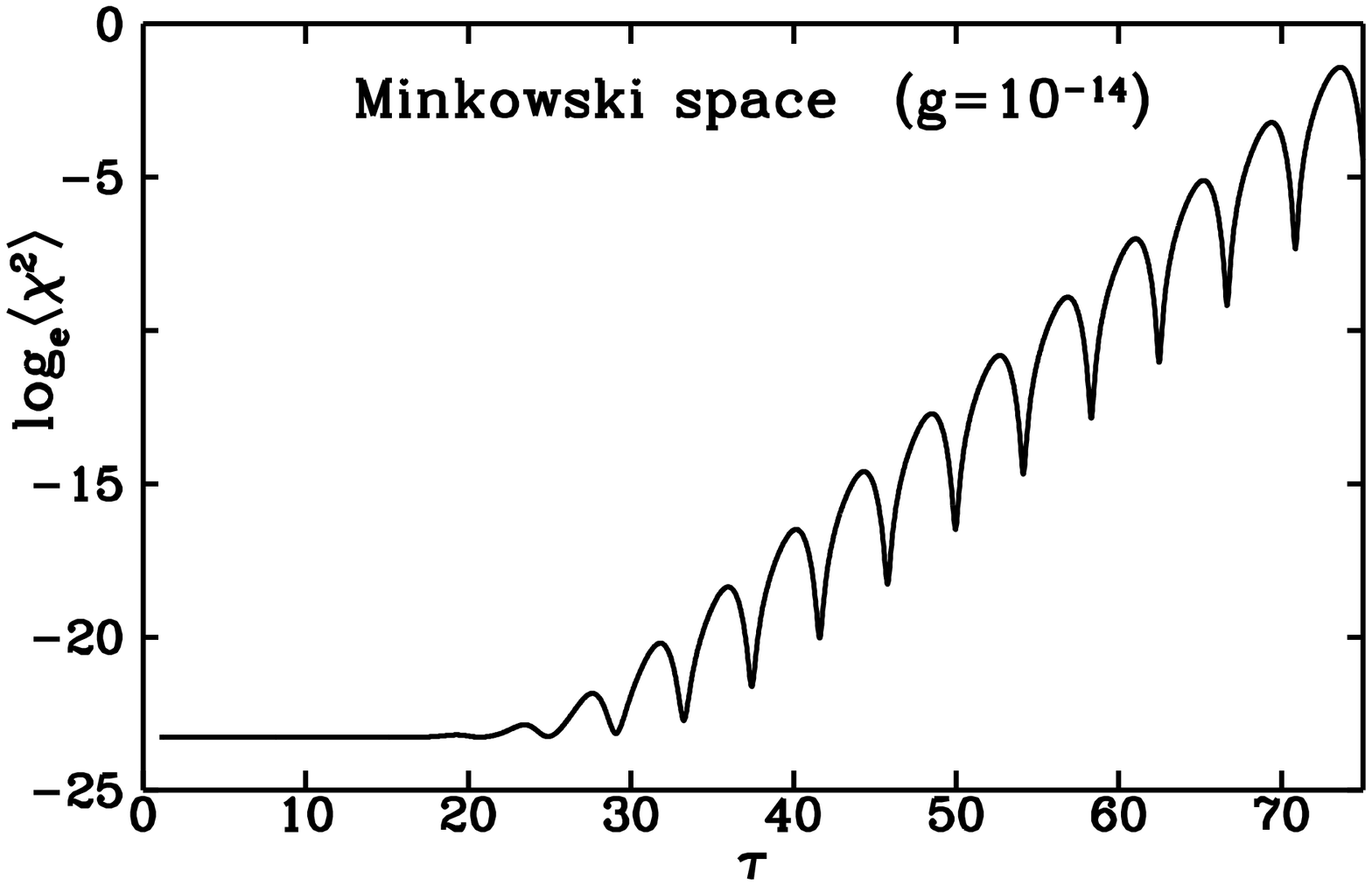} }
\footnotesize{Fig.\ 4b:  Time evolution of the supergravity and 
string theory  axion fluctuations in Minkowski space.}\\
\end{figure}

\end{document}